# Shallow geothermal energy potential for heating and cooling of buildings with regeneration under climate change scenarios

Alina Walch[1 *)], Xiang Li[2 *)], Jonathan Chambers[2], Nahid Mohajeri[3], Selin Yilmaz[2], Martin Patel[2], Jean-Louis Scartezzini[1]

[1] Solar Energy and Building Physics Laboratory, EPFL, Switzerland

[2] Chair for Energy Efficiency, University of Geneva, Switzerland

[3] Institute for Environmental Design and Engineering, University College London, UK

*) Equal contribution

## Abstract

Shallow ground-source heat pumps (GSHPs) are a promising technology for contributing to the decarbonisation of the energy sector. In heating-dominated climates, the combined use of GSHPs for both heating *and* cooling increases their technical potential, defined as the maximum energy that can be exchanged with the ground, as the re-injection of excess heat from space cooling leads to a seasonal regeneration of the ground. This paper proposes a new approach to quantify the technical potential of GSHPs, accounting for effects of seasonal regeneration, and to estimate the useful energy to supply building energy demands at regional scale. The useful energy is obtained for direct heat exchange and for district heating and cooling (DHC) under several scenarios for climate change and market penetration levels of cooling systems. The case study in western Switzerland suggests that seasonal regeneration allows for annual maximum heat extraction densities above 300 kWh/m$^2$ at heat injection densities above 330 kWh/m$^2$. Results also show that GSHPs may cover up to 55% of heating demand while covering 57% of service-sector cooling demand for individual GSHPs in 2050, which increases to around 85% with DHC. The regional-scale results may serve to inform decision making on strategic areas for installing GSHPs.

**Graphical abstract**

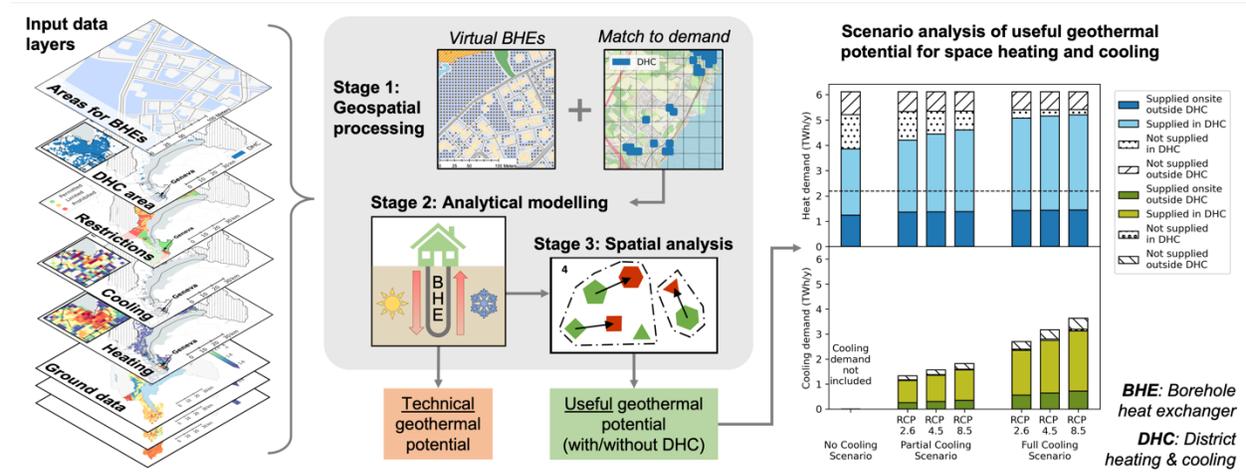

**Highlights**

- Technical potential of shallow ground-source heat pumps (GSHPs) at regional scale
- Seasonal regeneration of GSHPs by re-injection of space cooling needs to the ground
- Useful potential for supplying building energy demand under three climate scenarios
- Impact of district heating and cooling (DHC) on the useful potential of GSHPs
- Case study in western Switzerland: Up to 87% of energy potentially supplied for DHC

# Nomenclature

**Abbreviations**

| | |
|---|---|
| BHE | Borehole heat exchanger |
| CDD | Cooling degree days |
| COP | Coefficient of performance |
| D | Scenario with district heating and cooling |
| DH | District heating |
| DHC | District heating and cooling |
| DHS | District energy system |
| FC | Full cooling |
| GSHP | Ground-source heat pump |
| HDD | Heating degree days |
| HGSHP | Hybrid ground-source heat pump |
| HP | Heat pump |
| IPCC | The Intergovernmental Panel on Climate Change |
| LTDH | Low-temperature district heating |
| NC | Scenario of no cooling |
| ND | Scenario without district heating and cooling |
| PC | Scenario of partial cooling |
| RCP | Representative concentration pathway |
| TLM | Topographic landscape model |

**Variables**

| | |
|---|---|
| $B$ | Borehole heat exchanger spacing ($m$) |
| $COP_{cool}$ | Coefficient of performance for cooling |
| $COP_{heat}$ | Coefficient of performance for heating |
| $H$ | Borehole heat exchanger depth ($m$) |
| $HDD/CDD_{max}$ | Maximum monthly HDD/CDD |
| $H_{max}$ | Maximum allowed drilling depth ($m$) |
| $Q_{cool}$ | Useful geothermal potential for supplying cooling ($Wh$) |
| $Q_{extr}$ | Technical geothermal potential for heat extraction ($Wh$) |
| $Q_{field}$ | Annual extractable energy of a borehole field ($Wh$) |
| $Q_{heat}$ | Useful geothermal potential for supplying heat ($Wh$) |
| $Q_{inj}$ | Injected heat ($Wh$) |
| $N_B$ | Number of borehole heat exchangers in a field |
| $R_b^*$ | Borehole thermal resistance ($mK/W$) |
| $R_{field}$ | Mean thermal interference of all surrounding BHEs ($mK/W$) |
| $R_{LT}$ | Long-term resistance of the borehole ($mK/W$) |
| $R_{LT}^{'}$ | $R_{LT}$, weighted by annual operating time ($mK/W$) |
| $R_{seas}$ | Seasonal maximum thermal resistance ($mK/W$) |
| $R_{seas}^{'}$ | $R_{seas}$, weighted by maximum monthly operating time ($mK/W$) |
| $T_0$ | Ground surface temperature (°C) |
| $T_g$ | Ground temperature (°C) |
| $T_{mf}$ | Mean temperature of the heat carrier fluid (°C) |
| $T_{mf,c}$ | $T_{mf}$ in peak cooling mode (°C) |
| $T_{mf,h}$ | $T_{mf}$ in peak heating mode (°C) |
| $T_{mf,min}$ | Minimum temperature of the heat carrier fluid (°C) |
| $T_{mf,max}$ | Maximum temperature of the heat carrier fluid (°C) |
| $\alpha$ | Ground thermal diffusivity (*m²/s*) |
| $\delta T/\delta z$ | Temperature gradient in the ground (*K/m*) |
| $\lambda$ | Ground thermal conductivity (*W/(m·K)*) |
| $d_m$ | Number of days in each month |
| $q_{max}$ | Heat extraction power (*W/m*) |
| $q_{nom}$ | Nominal operating power (*W/m*) |
| $t_a$ | Number of hours in a year (8760 *h*) |
| $t_{dim}$ | Planning horizon (year) |
| $t_m$ | Number of hours in the month of maximum heating/cooling (*h*) |
| $t_{nom}$ | Nominal operating time in heating mode (*h*) |
| $t_{op,c}$ | Number of full-load cooling hours (*h*) |
| $t_{op,h}$ | Number of full-load heating hours (*h*) |
| $w_{hdd/cdd,max}$ | Weight attributed to maximum monthly operation |

# 1 Introduction

Shallow geothermal energy is a promising low-carbon source to meet heating and cooling demands of buildings. The most commonly used type of shallow geothermal system in many European countries, including Switzerland, are vertical ground-source heat pumps (GSHPs) [1]. These systems exchange heat with the ground through one or multiple borehole heat exchangers (BHEs) installed at depths of up to 400m [2]. As temperatures are rising and extreme heat events are becoming more frequent due to climate change, space cooling demands may increase worldwide by up to 750% in residential buildings and 275% in commercial buildings by 2050 [3]. In heating-dominated climates such as central Europe, growing cooling demand could motivate a combined use of shallow geothermal energy for heating and cooling of buildings, using the ground as seasonal heat storage [4]. The re-injection of excess heat from space cooling to the ground hereby permits its seasonal regeneration, which reduces negative impacts of geothermal installations on the surrounding shallow subsurface [5], [6]. This has two-fold benefits for the technical geothermal potential, defined as the maximum thermal energy that can be exchanged with the ground using GSHP technology. Firstly, it allows a renewable supply of cooling demand, and secondly, it increases the potential for heating.

Evaluating the potential of GSHPs with seasonal regeneration requires (i) determining the amount of excess heat available during the regeneration period, and (ii) linking the potential GSHP systems to buildings. While individual GSHPs are directly connected to a nearby building, district heating and cooling (DHC) systems allow to transport heat between areas with high geothermal potential and areas with high energy demand [7]. In particular, 4[th] generation DHC, also known as low-temperature district heating (LTDH), has become attractive in Europe due to improved system efficiency, ability to integrate renewable and low-temperature sources and low carbon emissions [8], [9]. DHC is thus a promising technology to increase the useful geothermal potential, defined as the useful energy for supplying building energy demands.

To date, many studies of the technical geothermal potential quantify the energy that may be extracted from a single installation at a given location, for example in Italy [10], [11], Spain [12], Chile [13] and southern Switzerland [14]. These studies focus on the quantification of ground parameters but neglect the impact of the built environment or other GSHP systems. The built environment and its impact on the available area for installing BHEs has so far been primarily considered in studies at district scale [15], [16]. However, these studies rarely account for thermal interference [17], referring to increased ground temperature changes around densely installed BHEs, which increases the environmental impact of GSHPs and reduces their technical potential [18]. Thermal interference is addressed in studies of hypothetical borehole fields [19], [20] or by studying thermal plumes around existing installations [21], [22]. The regional-scale effects of thermal interference and the available area for BHE installation on the technical GSHP potential have been considered in a previous study by the authors [23], which however did not consider seasonal regeneration.

Seasonal regeneration, defined as the re-injection of heat to the ground during summer, has been mentioned in city-scale studies as a possibility to increase the geothermal potential [24]. Different heat sources for seasonal regeneration of GSHPs, notably space cooling needs and solar thermal generation, are discussed and compared in [25]. Case studies of individual buildings with seasonally regenerated GSHPs, referred to as "hybrid GSHP" (HGSHP) in [4], can be found for buildings of the residential [26], service [27] or transport sector [28]. While these studies provide an indication of the potential of seasonal regeneration, the results are specific to each case study. A large-scale view of the regeneration potential is provided in some studies by comparing fictive HGSHP systems in across a number of locations, for example for 19 cities in the EU [29], 40 cities in Greece [30] or three locations in Australia [31]. A sensitivity analysis of five types of HGSHPs across several building types in North America is provided in [32]. None of these studies, however, quantify the impact of seasonal regeneration on the geothermal potential for an entire region.

Accounting for seasonal regeneration of GSHP systems requires matching the technical geothermal potential with building energy demands. Such matching has been done at building level [33], district level [17], and at city/large scale [34], [35]. However, these studies have only dealt with the mapping of potential installations to nearby buildings, thus not considering the potential of district energy systems (DHS). To date, the potential of shallow geothermal energy to supply DHS has mostly been assessed for case studies of individual DHS, often focusing on the techno-economic analysis of the DHS design [36], [37]. Some case studies also address the potential of combining geothermal energy with other renewables such as wind [38] and/or solar thermal energy [39], [40] in DHS. The design of DHS with geothermal heat sources has also been assessed at city scale, focusing on network design rather than technical limitations of geothermal systems [41]. At regional or national scale, a spatial mapping of potential heat sources for DHS has been provided for Denmark [42], but excludes geothermal resources. Stegnar et al. [43] have assessed the techno-economic potential of shallow geothermal energy for DHS in Slovenia, accounting for thermal interference and local ground characteristics. Of these studies, only Formhals et al. [40] consider seasonal regeneration, and no study beyond building scale quantifies the potential for supplying heating and cooling demands from GSHP systems.

To fill the above-mentioned gaps, this paper presents a novel framework to estimate the technical and useful shallow geothermal potential from GSHPs for space heating and cooling at regional scale. The proposed framework combines, for the first time, (i) the spatial mapping between building energy demands and potential GSHPs at regional scale, (ii) the analytical modelling of seasonal regeneration for GSHPs, and (iii) the optimization of heat supply for district heating and cooling (DHC). To this aim, we expand the analytical model for quantifying technical geothermal potential at regional scale, proposed by Walch et al. [23], to account for bi-directional GSHP operation (heat injection and heat extraction). In this work, excess heat from space cooling is considered for heat injection, but the proposed approach can also be used for other heat sources. We further apply a graph-theory based optimization for maximising the supply of technical geothermal potential to buildings using DHC. The method is applied to a case study in the cantons of Vaud and Geneva in Switzerland, the country with the world's highest density of direct geothermal energy use per land surface [1], mostly from GSHPs [44]. Following a scenario-based approach, we obtain the technical and useful shallow geothermal potential with and without DHC, for two market penetration levels of building cooling systems and for three climate change scenarios. To the best of our knowledge, the results present the first regional-scale estimate of shallow GSHP potential that combines space heating and cooling.

## 2   Methods

The proposed framework for assessing the technical and useful potential of GSHPs to cover building heating and cooling demands consists of three stages, as shown in Figure 1: First, geospatial processing is performed to match potential GSHP systems to building energy demands (Section 2.1, green boxes in Figure 1). Second, analytical modelling is used to quantify the technical heat exchange potential for GSHP systems (Section 2.2, blue boxes in Figure 1). The considered GSHP systems consist of vertical closed-loop BHEs, from which heat is extracted during the winter months for space heating and re-injected using excess heat from space cooling during the summer months, leading to seasonal regeneration. Third, spatial analysis is applied to optimally allocate the technical potential between individual GSHPs and DHC, yielding the useful geothermal potential (Section 2.3, orange box in Figure 1). All modelling steps described below were implemented using the *python* programming language.

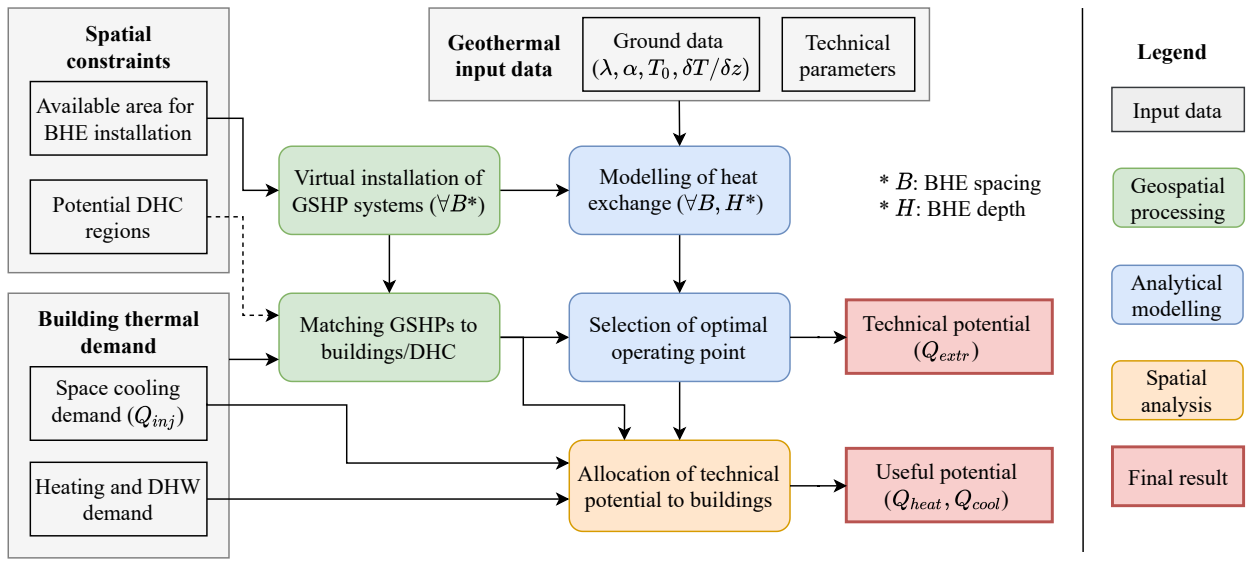

Figure 1: Workflow for modelling the technical potential of GSHP ($Q_{extr}$) and the useful potential for supplying heat ($Q_{heat}$) and cooling ($Q_{cool}$), consisting of three stages: (1) Geospatial processing (green boxes), (2) Analytical modelling (blue boxes), and (3) Spatial analysis (orange box), corresponding to the following sub-sections. Dashed lines are used only if DHC is considered.

## 2.1 Geospatial processing

To assess the potential for ground-source heat pumps at regional scale, we use a spatial mapping approach to (i) assign the location and arrangement (i.e. spacing) of potential BHE installations and (ii) match the resulting virtual borehole fields with the building energy demand.

*2.1.1 Virtual installation of GSHP systems*

To assign the location and arrangement of potential BHEs, we use the estimated available areas for BHE installation provided in [23], which have been derived from parcel data. These parcels represent individual property units, from which building footprints, other built-up areas (roads, railways, traffic-related areas and leisure zones) and natural habitat that is likely unsuitable for the installation of BHEs (water bodies, forests and wetlands, protected areas) have been removed. To obtain the available areas, a buffer of 3m has further been subtracted from parcel boundaries and building footprints, as specified in the technical norm for geothermal installations of the Swiss association of engineers and architects (SIA) [45].

On these available areas, we virtually install BHEs as rectangular grids with spacings ($B$) ranging from 5m to 100m, assuming that all available area is covered by these BHE grids. The lower boundary for $B$ (5m) equals the minimum distance between BHEs in a field as defined in the SIA norm [45], while the upper boundary (100m) equals half of the maximum considered borehole depth (see Section 3.2.2). As the thermal interference between boreholes decreases logarithmically with borehole spacing [23], the simulated $B$ follow this pattern ($B \in \{5, 7, 10, 15, 20, 25, 30, 40, 50, 70, 100\}\ m$). The intersection of the BHE grids with the available area yields the individual GSHP systems, one for each parcel and for each selected spacing $B$.

*2.1.2 Matching of GSHPs to building energy demand*

To match the potential GSHP systems to building energy demands, spatial constraints for the transportation of heat must be considered. In this step, we virtually connect GSHPs to buildings, for individual GSHP systems and for DHC. Generally, individual GSHPs are only connected to buildings in the same parcel, subject to land ownership. To ensure the scalability of the approach to regional scale, the matching is done for pixels of 400×400m$^2$ resolution. If DHC does not exist, we hence assume that GSHPs and buildings inside the same pixel are connected. By contrast, DHC enables the connection of buildings and GSHPs with a distribution network, such that all buildings are reachable by thermal energy from GSHPs. Consequently, in areas where DHC exists, we match all GSHPs and buildings inside the same DHC and treat them as an interconnected system.

As heating demands exceed cooling demands in heating-dominated climates, the analytical model described in the following is designed to estimate the maximum heat extraction for a given amount of injected heat. Thus, the level of heat injection, here given by the space cooling demand, must be assigned to each parcel. For this, we rank all parcels within a given pixel or DHC based on their area. As the benefits of seasonal regeneration increase with the level of cooling re-injection, we re-inject the maximum possible amount of heat to the largest borehole field. If the cooling demand exceeds the maximum capacity of the largest field, cooling energy is re-injected into the second largest field and to the following parcels until all cooling demand is satisfied. This approach reduces the number of BHE fields that are used bi-directionally for heating and cooling, which is more realistic than a small heat re-injection in all fields.

## 2.2 Analytical modelling of heat exchange potential for bi-directional GSHPs

The proposed analytical model of bi-directional GSHPs expands upon previous work [23] to account for seasonal regeneration through the re-injection of heat to the BHEs. To this aim, a heat injection load ($Q_{inj}$) is added during the summer season, which reduces the long-term temperature drop in the ground. Consequently, the thermal interference between BHEs is reduced and the heat extraction potential ($Q_{extr}$) is increased. We further add technical limitations for heat injection and consider a higher operating time, which is typical for GSHPs with seasonal regeneration [28], [46]. We focus on space cooling as heat source, but the methodology can be equally used for other sources of heat injection, such as solar thermal or industrial waste heat.

The proposed method follows a two-step approach: First, we simulate the annual extractable energy of a borehole field ($Q_{field}$), the maximum heat extraction power ($q_{max}$), and the number of full-load heating hours ($t_{op,h}$) for a range of borehole spacings $B$ and depths $H$, as well as for two operating modes (nominal $t_{op,h}$ and nominal $q_{max}$), such as to comply with the installation standards defined in the SIA norm [45]. Second, the borehole arrangement of each GSHP system is optimised by choosing the $B$, $H$ and operating mode that maximise the technical potential ($Q_{extr} + Q_{inj}$) while sustaining a feasible $q_{max}$ and $t_{op,h}$. As the methodology is aimed at regional-scale potential analyses, it is assumed that the heat pumps of each GSHP system are well-sized to supply the estimated $Q_{extr}$, $Q_{inj}$ and $q_{max}$. For simplification, we further assume that all systems start their operation simultaneously at time $t = 0$.

### 2.2.1 Modelling of heat exchange potential

The annual extractable energy of a borehole field ($Q_{field}$), which is simulated for each borehole arrangement ($B, H$), is defined as [23]:

$$Q_{field} = q_{max} \times t_{op,h} \times H \times N_B \quad (1)$$

where $q_{max}$ is the maximum heat extraction power (in *W/m*) and $N_B$ is the number of BHEs in the field. While $H$ and $N_B$ are assumed to be fixed for a given simulation, $q_{max}$ and $t_{op,h}$ are free parameters that need to be selected. To assure feasible operating conditions of the GSHPs, we constrain $q_{max}$ to at least 80% of the nominal operating power [23] and $t_{op,h}$ between the nominal operating time $t_{nom}$ (residential heating only) [45] and the maximum operating time, which assumes that the GSHP is operated 100% of the time in the month with maximum heating load:

$$q_{max} \geq 80\% \, q_{nom} \quad (2)$$

$$t_{nom} \leq t_{op,h} \leq \frac{t_m}{w_{hdd,max}} \quad (3)$$

where $t_m$ is the number of hours in the month of maximum heating operation.

The choices of $q_{max}$ and $t_{op,h}$ are further constrained by the mean temperature of the heat carrier fluid inside the BHE ($T_{mf}$). The $T_{mf}$ must never drop below a minimum value $T_{mf,min}$ in heating mode and must not exceed a maximum value $T_{mf,max}$ during cooling operation. Both constraints must be fulfilled at all times, from the first year to the last year of the planning horizon ($t_{dim}$). To simulate the heat transfer between BHE fields and the ground, we use Eskilson's analytical model [47], which represents each BHE as a finite line source. Following the principles of spatial and temporal superposition, we model the BHE operation by superimposing a long-term and a seasonal heat extraction component, as well as the thermal interference of all surrounding boreholes. The $T_{mf}$ then equals the undisturbed ground temperature ($T_g$) at half the borehole depth minus the sum of the temperature drops due to each superimposed component. To not violate any temperature constraint, the chosen $q_{max}$ and $t_{op,h}$ must hence fulfil the following equations for peak heat extraction (heating mode) and heat injection (cooling mode) in the first ($t = 0$) and last year ($t = t_{dim}$) of operation (cf. [46], [48]):

$$T_{mf,min} \leq T_{mf,h}(t) = T_g - q_{max,h}\left(R'_{LT,h}(t) + R'_{seas,h} + R^*_b\right) + \frac{Q_{inj}}{N_B t_{op,c} H} R'_{LT,c} \quad (4)$$

$$T_{mf,max} \geq T_{mf,c}(t) = T_g + \frac{Q_{inj}}{N_B t_{op,c} H}\left(R'_{LT,c}(t) + R'_{seas,c} + R^*_b\right) - q_{max,h} R'_{LT,h} \quad (5)$$

where $T_{mf,h}$ and $T_{mf,c}$ are the $T_{mf}$ in peak heating ($h$) and cooling ($c$) modes; $Q_{inj}$ is the injected heat (in Wh); $t_{op,c}$ (in h) is the operating time in cooling mode; $R^*_b$ is the borehole thermal resistance (in mK/W); $R'_{LT}$ and $R'_{seas}$ denote the long-term and seasonal ground thermal resistance, weighted for heating or cooling operation. At $t = 0$, we assume that $R'_{LT} = 0$. At $t = t_{dim}$, the $R'_{LT}$ is weighted for annual mean operation, given by the fraction of operating time, and is composed of the long-term resistance of the borehole itself ($R_{LT}$) and the mean thermal resistance of all other BHEs within and around a field ($R_{field}$), which depends on $B$ and $H$:

$$R'_{LT,h/c}(t) = \begin{cases} 0 & t = 0 \\ \frac{t_{op,h/c}}{t_a}\left(R_{LT}(H) + R_{field}(B, H) - R_{seas}\right) & t = t_{dim} \end{cases} \quad (6)$$

where $t_a = 8760h$ and $R_{seas}$ is the seasonal maximum thermal resistance, which is subtracted for mathematical consistency of the model. Following [23], we model $R_{LT}$ and $R_{field}$ from a heat extraction pulse of duration $t_{dim}$, and $R_{seas}$ from the peak of a sinusoidal heat extraction with periodicity of 1 year. The thermal resistances are functions of the ground parameters, which vary regionally (see Section 3.2.2), and the borehole geometry. $R_{field}$ further depends on the borehole arrangement within the field, decreasing logarithmically as $B$ increases. The mathematical formulations for $R_{field}$, $R_{LT}$ and $R_{seas}$ are provided in [23].

The $R'_{seas}$ is the maximum monthly thermal resistance, given by multiplying $R_{seas}$ with the fraction of the maximum monthly operating time:

$$R'_{seas,h/c} = \frac{w_{hdd/cdd,max} \, t_{op,h/c}}{t_m} R_{seas} \quad (7)$$

where $t_m$ is the number of hours in the month of maximum heating/cooling operation, and $w_{hdd/cdd,max}$ is the weight attributed to maximum monthly operation. The weight is obtained from the heating degree days (HDD) for heating mode and from the cooling degree days (CDD) in cooling mode and allows to account for the monthly variation of the heating and cooling demand (see Appendix A). Due to the lack of a norm for cooling operation, we set $t_{op,c}$ to the maximum operating time, such that $t_{op,c} = t_m/w_{cdd,max}$ in analogy to Eq. (3).

*2.2.2 Selection of optimal operating point*

To select the optimal operating point of a GSHP system with a given $B$ and $H$ for any level of $Q_{inj}$, we aim to find the highest $q_{max}$ and $t_{op,h}$ that fulfil Eqs. (2)-(5) at any time. To reduce the complexity of the selection, we consider two operating modes. The first represents a heating-only use, where $t_{op,h}$ equals the nominal value ($t_{nom}$, 1800-2000h in Switzerland) [45]. We thus compute the highest $q_{max}$ that fulfils all constraints by fixing $t_{op,h} = t_{nom}$. The second operating mode represents large installations used for both heating and cooling. These systems typically have higher operating times, around 2500-3000h [46], while being operated at nominal power ($q_{nom}$). We hence fix $q_{max} = q_{nom}$ and maximise $t_{op,h}$. This second configuration is often infeasible for heating-only cases, but it is suitable for seasonal regeneration because the long-term temperature drops around BHEs decreases as more heat is injected.

Using Eq. (1), $Q_{field}$ is then computed for each $B$, $H$ and operating mode. Out of these, the heat extraction potential of the borehole field ($Q_{extr}$) and the optimised borehole arrangement are obtained as the feasible solution that maximises the heat exchange potential:

$$Q_{extr} = \max_{B,H,op.mode} Q_{field} + Q_{inj} \quad \text{subject to Eqs. (2)} - (5) \; \forall t, H \leq H_{max} \quad (8)$$

where $H_{max}$ represents the maximum allowed drilling depth (see Section 3.2.1).

To obtain the heating ($Q_{heat}$) and cooling ($Q_{cool}$) energy exchanged with the buildings from $Q_{extr}$ and $Q_{inj}$, the coefficient of performance (COP) of the heat pumps (HP) must be taken into account. We model the GSHPs as water-to-water heat pumps, which are prevalent in DHC [49]. Expecting a small increase in future HP performance compared to current COPs [50], [51], we choose a constant COP for heating ($COP_{heat}$) as 4.5 and for cooling ($COP_{cool}$) as 5.5, such that (cf. [50]):

$$Q_{heat} = Q_{extr} \frac{COP_{heat}}{(COP_{heat} - 1)} \tag{9}$$

$$Q_{cool} = Q_{inj} \frac{COP_{cool}}{(COP_{cool} + 1)} \tag{10}$$

## 2.3 Spatial analysis of useful geothermal potential

Once the technical potential is modelled, we map it with building heating demand on-site, based on matching results derived in Section 2.1.2. The useful potential for heating and cooling is defined as the portion of technical potential that is smaller than or equal to the heat demand of matched buildings. The portion of technical potential that exceeds the heat demand of matched buildings is considered as surplus potential. The surplus potential is only obtained for heating, as the analytical model described in Section 2.2 always attempts to inject the maximum possible amount of cooling demand. There are also areas where technical potential is insufficient to supply building heating or cooling, which is considered as deficit.

If DHC is considered for heat distribution, we further allocate surplus potential to nearby DHC areas with a deficit, following the on-site mapping. This allows an increase in the useful potential, as DHC is usually located in dense areas with a high demand. These areas likely have insufficient technical potential to supply the demand. The allocation process is adapted from the method of Chambers et al. [52]. This allows to analyse the supply of DHC with geothermal energy in neighbouring areas, accounting for the spatial constraint that the potential energy sources need to be within a limited range of DHC. The key strength of this method is that it uses spatial analysis and graph theory to disaggregate a large-scale case into sub-clusters of supply and demand, thus expanding the applicable spatial scale. However, this method allocated energy using the simple net-balance across each sub-cluster, which assumes that energy can be allocated from supplies to demands that are not directly connected. This limitation is addressed in the present study by introducing an optimisation algorithm which finds the maximum allocation of supply to demand within each sub-cluster. A detailed description of the new method is presented in a methods paper accompanying this research article [53]. Figure 2 shows an overview of the method to link surplus geothermal potential and DHC demand. It consists of the following key steps:

Step 1. Each area with surplus geothermal potential and each DHC area with a deficit are treated as the vertices of a bipartite graph. These vertices are characterized by their geometries and heating capacities (either supply surplus or deficit). We call vertices representing areas with surplus geothermal potential 'sources', and vertices representing DHC areas with deficit 'demand'.

Step 2. For each demand, all sources for which the closest distance to the demand is below a certain threshold are connected to the demand by an edge. To account for the variation in DHC sizes, the threshold is chosen as 20% of the length of the oriented minimum bounding box of the DHC area.

Step 3. The source/demand vertices and linking edges is converted into a large bipartite graph covering the whole study area, which is split into subgraphs of connected components (defined as a subset of vertices that are connected to each other, but not to any vertices outside the set).

Step 4. The maximum possible allocation of technical potential from sources to demands within each subgraph is formulated as a Hitchcock transportation problem [54] and then solved using linear programming [55].

Step 5. Finally, the maximum possible allocation of all subgraphs is summed and added to the usable potential estimated.

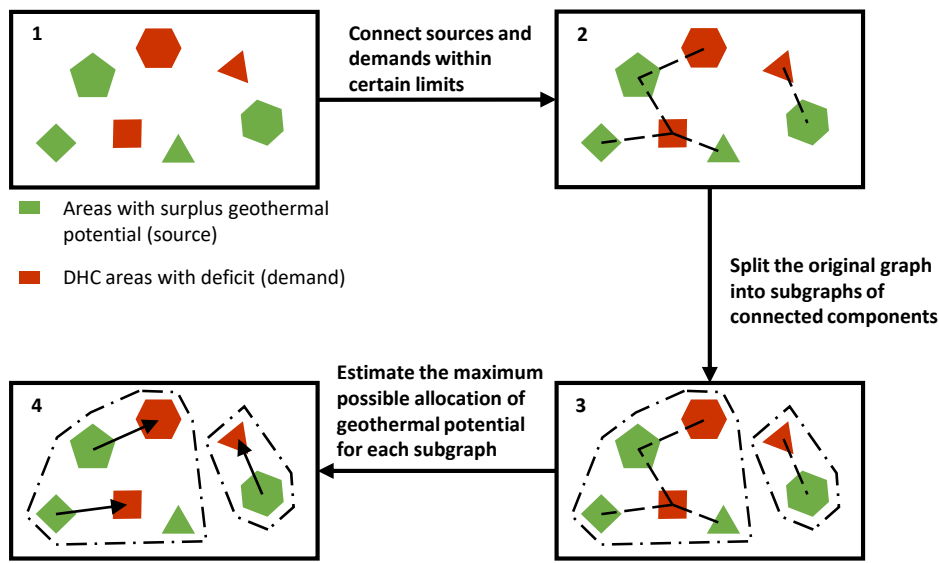

Figure 2. Overview of the method to allocate surplus geothermal potential to district heating and cooling areas.

## 3   Case Study

The proposed method is applied to a case study in the cantons of Vaud and Geneva in Switzerland. The case study area covers a surface of around 1,600 km$^2$ in western Switzerland, containing two of Switzerland's largest cities (Geneva and Lausanne). The presence of high-resolution geothermal cadastres and landscape data, as well as a rapid projected growth in cooling demand [56], [57], make this area particularly suitable for a regional-scale study of the shallow geothermal potential.

To match the technical geothermal potential with the building energy demand, we use simulated annual heating and service-sector cooling demands for the year 2050. This modelling horizon is chosen as (i) the adoption of space cooling technology is expected to increase in the coming years, and (ii) the installation of a large amount of GSHP systems would take several years. Furthermore, considering different climate change scenarios for 2050 allows to assess the robustness of the technical geothermal potential with seasonal regeneration in relation to climate change.

### 3.1   Scenarios considered

This work uses a scenario-based approach to assess the impact of seasonal regeneration from space cooling and the existence of DHC on the useful geothermal potential. Table 1 presents the different scenario components, namely three levels of space cooling demand, three climate change models and the possibility to use DHC. The levels of space cooling demand correspond to different penetration levels of cooling equipment in the building stock. These levels are (i) a reference case without cooling ('no cooling'), (ii) the projected penetration of cooling equipment under current tight regulation ('partial cooling'), and (iii) an extreme case saturating nearly all cooling demand ('full cooling'). The climate change models describe three representative concentration pathways (RCPs) adopted by the IPCC [58]. These are linked to the space cooling demand by modelling the diffusion of space cooling devices and the cooling intensity corresponding to each RCP. To assess the robustness of the potential estimate to climate change, we choose a stringent emission reduction scenario (RCP 2.6), an intermediate scenario (RCP 4.5), and a worst-case scenario (RCP 8.5). The method presented in Section 2 is applied to each combination. All scenario combinations and their naming convention used throughout the paper are shown in Table 2.

Table 1. Description of scenario components.

| Scenario components | Levels | Description |
|---|---|---|
| Space cooling demand | No Cooling | No space cooling demand met by GSHP (no seasonal regeneration). |
| | Partial Cooling | Only a portion of buildings are actively cooled. The diffusion of space cooling equipment and the growing space cooling demand is predicted based on the past trend. |
| | Full cooling | Extreme scenario where space cooling demand is mature and nearly saturated. |
| IPCC climate change models | RCP 2.6 | Stringent emission reduction scenario where emissions peak around 2020 [58]. |
| | RCP 4.5 | Intermediate scenario where emissions peak around 2040 [58]. |
| | RCP 8.5 | Worst-case scenario where emissions increase throughout the 21st century [58]. |
| DHC utilization | Without DHC | No DHC is installed (GSHPs are only connected to on-site building thermal demands). |
| | With DHC | In each DHC potential area, the distribution network connects all GSHPs and buildings. This allows integrating surplus geothermal potential within a limited range. |

Table 2. Summary of scenario combinations

| | *Climate model* | **Without DHC (ND)** | **With DHC (D)** |
|---|---|---|---|
| No Cooling (NC) | | Base scenario (NC-ND) | NC-D |
| Partial Cooling (PC) | *RCP 2.6* | PC-ND-2.6 | PC-D-2.6 |
| | *RCP 4.5* | PC-ND-4.5 | PC-D-4.5 |
| | *RCP 8.5* | PC-ND-8.5 | PC-D-8.5 |
| Full Cooling (FC) | *RCP 2.6* | FC-ND-2.6 | FC-D-2.6 |
| | *RCP 4.5* | FC-ND-4.5 | FC-D-4.5 |
| | *RCP 8.5* | FC-ND-8.5 | FC-D-8.5 |

## 3.2 Regional datasets

Conducting a regional-scale study of geothermal potential requires the availability of high-quality data on the ground thermal properties, the building energy demand and the spatial constraints for the installation of GSHPs and the coverage of the energy demand (see Figure 1). Table 3 provides an overview of all regional datasets and their sources.

Table 3. Overview of regional datasets

| | Dataset | Description | Resolution | Sources |
|---|---|---|---|---|
| Spatial constraints | Parcel boundaries | Boundaries of public & private property units | Polygons | ASIT-VD [59], SITG [60] |
| | Topographic Landscape Model | Incl. building footprints, other built-up areas, natural habitat | Polygons | SwissTopo [61] |
| | DHC zones | Potential areas for DHC | Polygons | Chambers et al. [52] |
| | GSHP restrictions | Permitted, limited and prohibited zones for GSHPs | Polygons | ASIT-VD [62], SITG [63] |
| Geothermal input data | Thermal ground properties | Thermal conductivity & diffusivity | $50 \times 50 \times 50 m^3$ (50-300m depth) | ASIT-VD [62], SITG [63] |
| | Surface temperature | Average ground surface temperature at 1m depth | $200 \times 200 m^2$ | Assouline et al. [64] |
| | Air temperature | Daily mean air temperature at 2m above ground | $1 \times 1 km^2$, daily (1991-2010) | MeteoSwiss [65] |
| | Digital Elevation Model | Elevation map | $2 \times 2 m^2$ | SwissTopo [66] |
| Building energy demands | Building heating demand | Space heating and domestic hot water demand | Buildings, annual | Chambers et al. [67], Schneider et al. [68] |
| | Building cooling demand | Building space cooling demand in the service sector | Buildings (service sector), annual | Li et al. [69] |

*3.2.1 Spatial constraints*

Spatial constraints for the mapping of shallow GSHP potential include (i) the parcels, buildings and landscape features, required to estimate the available area for borehole installation (Section 2.1.1), (ii) potential areas for DHC (Section 2.3), and (iii) restrictions for GSHP installation (see below). The parcel boundaries are based on the official mensuration data for roughly 100,000 property units. The topographic landscape model (TLM) contains a 3D representation of various landscape objects, some of which are unsuitable for installing BHEs. A 1m buffer is added around all unsuitable objects to account for inaccuracies of the TLM. Potential areas for low temperature district heating networks, shown in Figure 3a, are obtained from an existing model [52]. Restrictions on the installation of GSHPs are divided into permitted, limited and prohibited zones (Figure 3b). In permitted zones we assume a maximum drilling depth ($H_{max}$) of 200 m, in limited zones of 150 m, while no GSHP systems are installed in prohibited zones. These values are obtained based on typical drilling depths of existing installations in each zone, as shown in [23].

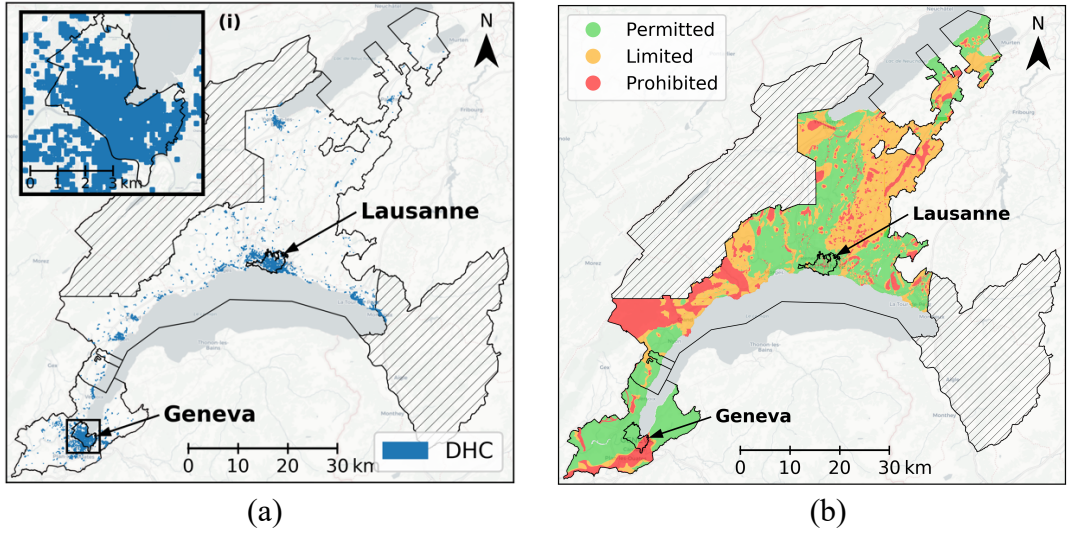

Figure 3. Spatial constraints, a) for the potential DHC areas, and b) for the permitted areas of BHE installation. Hatched areas are not included in the case study due to a lack of available ground data.

### 3.2.2 Geothermal input data

Estimating the technical geothermal potential (Section 2.2) requires an estimate of the long-term and seasonal thermal resistance ($R_{LT}$, $R_{seas}$), the nominal operating time ($t_{op,h}$), the seasonal load ($w_{hdd/cdd,max}$) and the nominal heat extraction rate ($q_{nom}$) for each borehole configuration. The borehole configurations are discretised for each parcel into 11 borehole spacings (see Section 2.1.1) and four depths, ranging from $50m$ to $H_{max} = 200m$ following the spatial resolution of the ground data (Table 3). The thermal resistance for a dimensioning horizon of $t_{dim} = 50$ years is obtained from previous work [23], where it was derived from regional-scale data of the ground thermal conductivity ($\lambda$), diffusivity ($\alpha$), the surface temperature ($T_0$) and the temperature gradient in the ground ($\delta T/\delta z$), which may be approximated as 0.03 K/m in the case study region [45]. The ground data is equally used to compute $q_{nom}$ and $T_g$ for each $H$ based on the guidelines in the SIA norm [45]. The $t_{nom}$ is mapped from the altitude following [45] based on a digital elevation model, assuming that it corresponds to the minimum operating time for all building types (residential and service sector). The degree days used to estimate $w_{hdd/cdd,max}$ are derived from gridded data of daily mean ambient temperature, averaged across 20 years [70]. As the data has a coarse resolution of 1×1km2, we spatially interpolate it to a grid of 200×200m2 pixels. Further technical parameters are the minimum and maximum fluid temperatures, which are set to $T_{mf,min} = -1.5°C$ and $T_{mf,max} = 50°C$. These temperatures limits are chosen based on the SIA norm [45] (for $T_{mf,min}$) and existing installation examples in Switzerland [46] (for $T_{mf,max}$) such as to avoid damage of the heat exchanger tubes due to freezing or overheating.

*3.2.3 Building thermal demands*

An existing model was used to estimate total demand per building for heating and hot water on a yearly basis [67], [68]. This is a regression-based model where typical measured heat demand intensities were linked to different building types based on the extensive metadata included in the Swiss Building Registry. A 50% reduction of demand is then applied uniformly to all building, based on the target of the Swiss Energy Strategy 2050 [71]. This 50% reduction heat demand scenario is used consistently across all scenarios, yielding the energy demand shown in Figure 4a. In total of all buildings in the studied area, the heating demand is 6.11 TWh/y.

Building cooling demand for the partial cooling (PC) and the full cooling (FC) scenarios are generated using the Monte Carlo model introduced in the work [69]. The Monte Carlo model forecasts future building cooling demand in the service sector (e.g. offices, trade, hotels, etc.) by applying a probability distribution for the adoption of space cooling equipment in buildings, to thereby estimate the magnitude and uncertainty in the cooling demand as a function of current and future building characteristics and climate. The results are reported in Table 4 for 2000 iterations of the model. An example for a single scenario and Monte Carlo run is shown in Figure 4b.

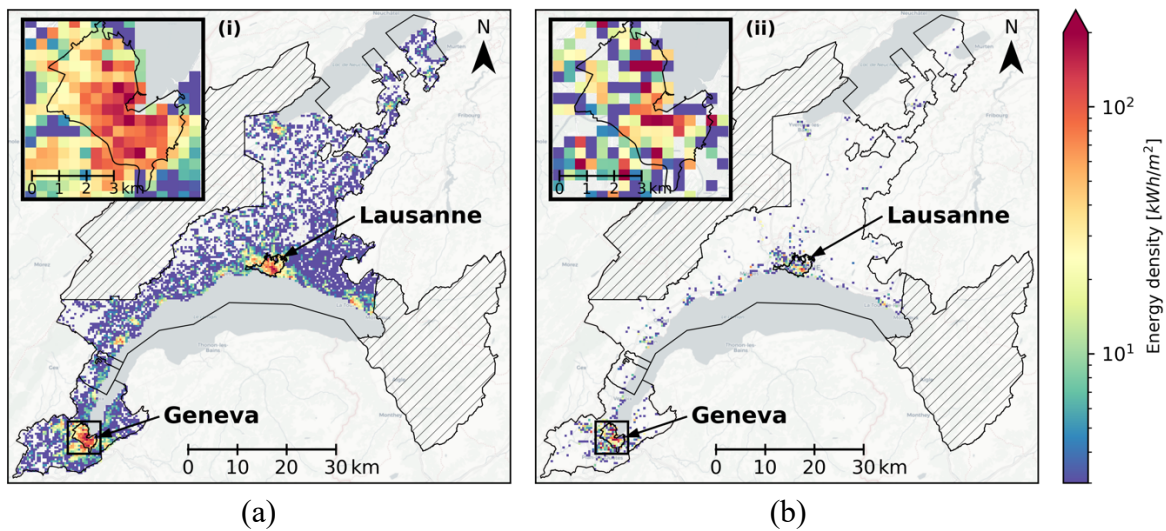

Figure 4. Building energy demand density, a) for space heating and domestic hot water and b) for space cooling for one Monte-Carlo simulation (climate change model RCP 4.5, partial cooling). The insets (i), (ii) show the city of Geneva.

Table 4. Mean and 95% confidence interval of total building cooling demand in all scenarios

|  | Climate model | Total building cooling demand (TWh/y) |
|---|---|---|
| No Cooling (NC) |  | 0 |
| Partial Cooling (PC) | RCP 2.6 | 1.33±0.18 |
|  | RCP 4.5 | 1.57±0.21 |
|  | RCP 8.5 | 1.82±0.24 |
| Full Cooling (FC) | RCP 2.6 | 2.70±0.18 |
|  | RCP 4.5 | 3.17±0.21 |
|  | RCP 8.5 | 3.63±0.24 |

# 4 Results
## 4.1 Impact of cooling injection on technical geothermal potential

The case study in western Switzerland shows that seasonal regeneration from the re-injection of space cooling demands increases the technical heat extraction potential ($Q_{extr}$) significantly. While the scenario without regeneration (NC-ND) shows a maximum annual energy density of around 15 kWh/m² per pixel of 400×400m² (Figure 5a), the maximum heat extraction density exceeds 300 kWh/m² in pixels with high levels of heat injection (> 330 kWh/m²), as shown in Figure 5b and c. This 20-fold increase is explained by the strongly reduced thermal interference between boreholes, which has two-fold effects on the technical potential. Firstly, the number of boreholes increases quadratically as the average BHE spacing is reduced from $B_{opt} = 20-25m$ to $B_{opt} = 5-7m$ (Figure 6a). Secondly, reduced thermal interference allows for higher operating power and time (Figure 6b and c), which increases the heat extraction per borehole. These high energy densities are found primarily in dense urban areas, such as the city centre of Geneva (see insets in Figure 5 and Figure 6).

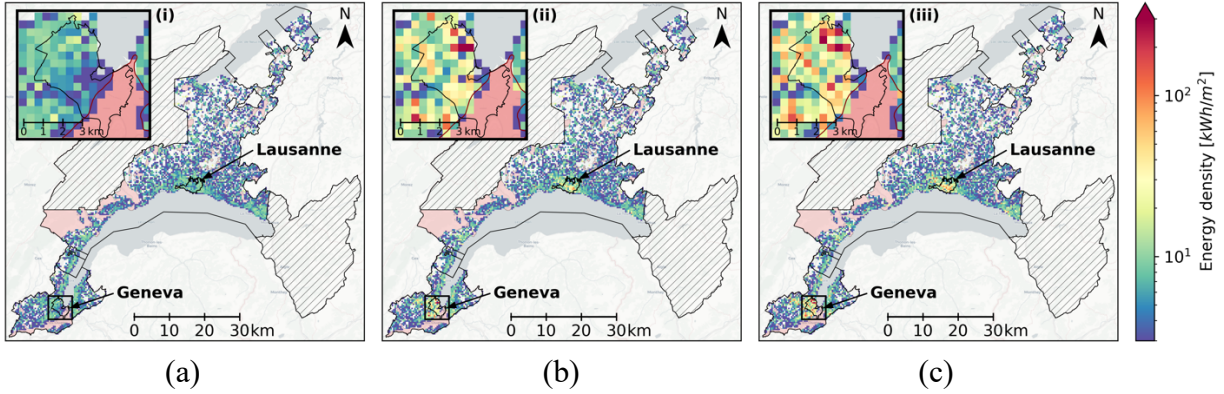

Figure 5: Heat extraction potential ($Q_{opt}$) of individual GSHPs, aggregated to pixels of 400×400m², for (a) baseline scenario (NC-ND), (b) partial cooling (PC-ND-4.5) and (c) full cooling scenario (FC-ND-4.5). In pink zones, GSHPs installation is prohibited.

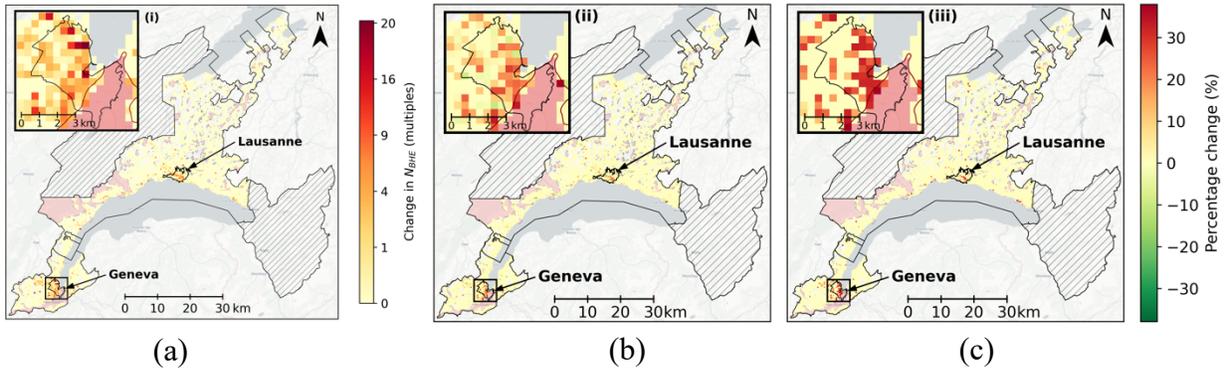

Figure 6: Change in number of boreholes $N_{BHE}$ (a), heat extraction rate $q_{max}$ (b) and heating operating time $t_{op,h}$ (c) of the partial cooling scenario (PC-ND-4.5) compared to the baseline (NC-ND), computed as the difference between the two scenarios divided by the baseline. The change in $N_{BHE}$ is shown as multiples on a quadratic scale.

In absolute terms, the technical potential can be increased by around 1 TWh if no DHC is considered (PC-ND), by 1.5 TWh with DHC (PC-D) for partial cooling, and by 2-3 TWh for full cooling (FC) (see $Q_{extr}$ in Table 5). The results are relatively robust to different climate change scenarios, varying by ±5% with respect to the RCP 4.5 climate model. The confidence intervals of the Monte Carlo runs vary by 5-10% for heat injection and <5% for extraction, suggesting that the total heat exchange is equally robust to the spatial distribution of cooling demand within the studied area. The scenarios further provide insights into the relation between the injected and extracted heat. As the linear fit of all scenarios (Figure 7) shows, roughly 90% of the injected heat during summer can be extracted in winter in addition to the baseline potential of 4.6 TWh. This high conversion rate of 90% is due to the strong effect of seasonal regeneration on reducing thermal interference between boreholes. These results show that seasonal regeneration is essential for a sustainable large-scale deployment of GSHPs.

Table 5: Mean and 95% confidence intervals of the technical heat exchange potential (heat injection and heat extraction) summed over the case study area for all scenarios, based on Monte Carlo simulation.

| Scenario | Heat injection $Q_{inj}$ (TWh/y) | | | Heat extraction $Q_{extr}$ (TWh/y) | | |
|---|---|---|---|---|---|---|
| Base scenario | 0 | | | 4.64 | | |
| PC-ND | 0.99 ± 0.10 | 1.15 ± 0.12 | 1.30 ± 0.13 | 5.56 ± 0.09 | 5.70 ± 0.10 | 5.84 ± 0.12 |
| FC-ND | 1.93 ± 0.15 | 2.21 ± 0.17 | 2.46 ± 0.34 | 6.42 ± 0.13 | 6.67 ± 0.15 | 6.89 ± 0.38 |
| PC-D | 1.35 ± 0.14 | 1.60 ± 0.17 | 1.85 ± 0.19 | 5.90 ± 0.13 | 6.12 ± 0.15 | 6.34 ± 0.17 |
| FC-D | 2.78 ± 0.23 | 3.24 ± 0.27 | 3.69 ± 0.30 | 7.18 ± 0.20 | 7.59 ± 0.23 | 7.98 ± 0.26 |
| *Climate model* | *RCP 2.6* | *RCP 4.5* | *RCP 8.5* | *RCP 2.6* | *RCP 4.5* | *RCP 8.5* |

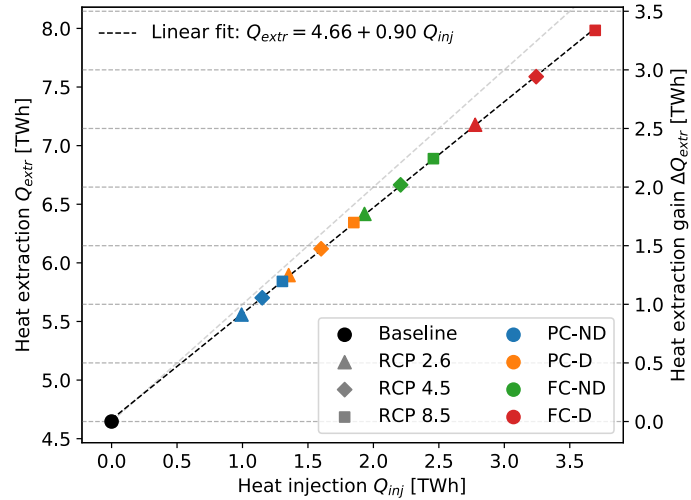

Figure 7: Geothermal potential ($Q_{extr}$) as a function of injected excess heat ($Q_{inj}$), for all scenarios (see Table 2). The black dashed line shows the linear fit, the grey dashed line represents a slope of 1.

### 4.2 Supply of building heating and cooling demands with geothermal potential

The useful potential to supply heating and cooling demand in seven scenarios of cooling penetration and climate change without DHC is presented in Table 6 and Figure 8. Due to the spatial constraint that geothermal energy can only supply demand in a limited range from the GSHP, only a fraction of the technical potential is useful. In the base scenario (NC-ND), only 2.2 TWh of the 6.0 TWh of $Q_{heat}$ (from 4.6 TWh of $Q_{extr}$ provided to HPs) is useful, which would cover 35% of building heating demand. When considering seasonal regeneration from space cooling in the PC-ND scenario, 0.8-1.1 TWh of space cooling demand could be supplied by GSHPs while increasing the useful potential for supplying heat demand by 0.7-0.8 TWh. Similarly, in the FC-ND scenario, GSHPs could meet 1.6-2.0 TWh building cooling demand, as well as supply additional 1.1-1.2 TWh to building heating demand. The fraction of demand covered (values in brackets) is nearly constant across the three climate models for all scenarios, which suggests a high robustness of the results to climate change.

Figure 9 shows maps of the heating and cooling supply in percentage for the PC-ND-4.5 scenario as an example. In the case of supplying cooling (Figure 9a), we found that building cooling demand could be sufficiently supplied where GSHP installation is allowed, due to the high capacity of BHEs to inject excess heat from space cooling. By contrast, no cooling demand can be supplied in areas where GSHP installation is prohibited (see insets in Figure 9a). In the case of supplying heat (Figure 9b), despite an increased potential due to seasonal regeneration, heat demand is not sufficiently supplied in dense urban areas with an excessive demand density. Although the technical potential is abundant in low-density areas, it is not available to other areas with a deficit, due to spatial constraints. In addition, constraints on GSHP installation further limit the heat supply (see Figure 3b).

Table 6. Useful potential to supply building heating and cooling demands in scenarios without DHC.

| Scenario | Useful cooling potential $Q_{cool}$ (TWh/y) | | | Useful heating potential $Q_{heat}$ (TWh/y) | | |
|---|---|---|---|---|---|---|
| Base scenario | 0 | | | 2.19 (35%) | | |
| PC-ND | 0.84 (63%) | 0.97 (62%) | 1.10 (60%) | 2.90 (47%) | 2.96 (48%) | 3.02 (49%) |
| FC-ND | 1.63 (60%) | 1.87 (59%) | 2.08 (57%) | 3.26 (53%) | 3.32 (54%) | 3.37 (55%) |
| *Climate model* | *RCP 2.6* | *RCP 4.5* | *RCP 8.5* | *RCP 2.6* | *RCP 4.5* | *RCP 8.5* |

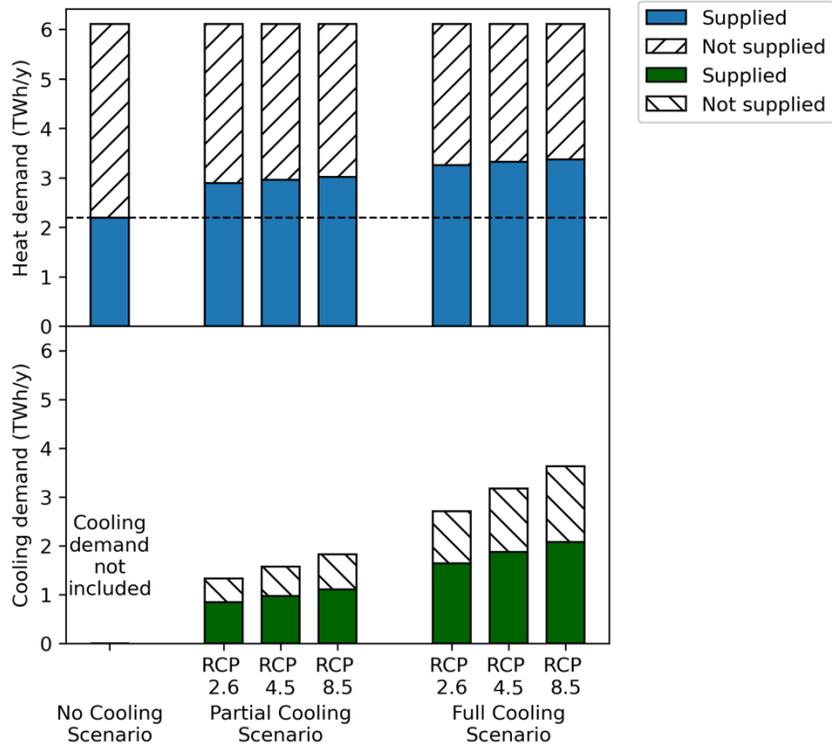

Figure 8. Supply of building heating and cooling demands in scenarios without DHC.

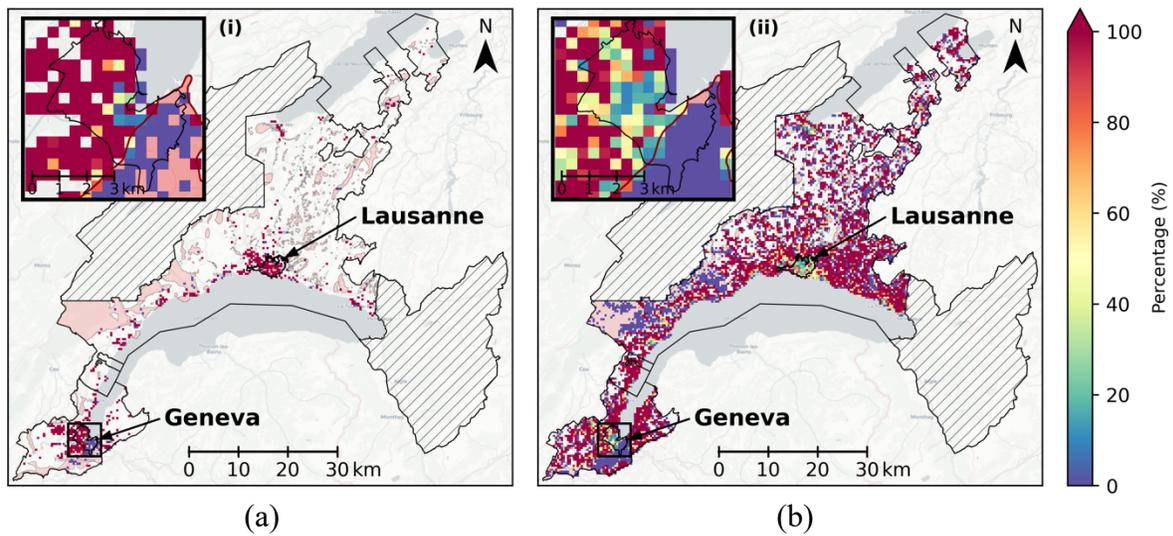

Figure 9. Distribution of demand coverage per pixel in the PC-ND-4.5 scenario: (a) percentage of cooling demand supplied; (b) percentage of heat demand supplied.

## 4.3 Impact of district heating and cooling

The integration of GSHPs in DHC increases the useful potential to supply heating and cooling demand by at least 23% percentage points compared to the results presented in Section 4.2, and allows to cover up to around 85% of building energy demands. As shown in Table 7, the utilization of DHC improves the useful potential in all seven scenarios with DHC, which are again robust to the climate models. The amount of energy supplied within DHC increases as more heat is re-injected to the ground, due to the high demand in dense areas (see Figure 10). These results demonstrate the two-fold benefit of DHC: First, more injection of space cooling demand results in an increased technical potential. Second, DHC eases the spatial constraints for the useful potential for heating and allows integrating more surplus geothermal energy in neighbouring areas as potential supply sources of the DHC.

To illustrate the impact of DHC at regional scale, Figure 11 shows maps of the change in heating and cooling supply between the PC-ND-4.5 and PC-D-4.5 scenarios as an example. In the case of cooling (Figure 11a), the utilization of DHC allows the distribution of cooling energy from GSHPs to buildings located in areas where GSHP installation is prohibited (see inset (i) in Figure 11a). In the case of heat supply (Figure 11b), DHC could largely improve the insufficient supply in dense urban areas.

Table 7. Useful potential to supply building heating and cooling demands in scenarios with DHC

| Scenarios | Useful cooling potential $Q_{cool}$ (TWh/y) | | | Useful heat potential $Q_{heat}$ (TWh/y) | | |
|---|---|---|---|---|---|---|
| NC-D | | 0 | | | 3.86 (63%) | |
| PC-D | 1.14 (86%) | 1.35 (86%) | 1.57 (86%) | 4.20 (69%) | 4.44 (73%) | 4.60 (75%) |
| FC-D | 2.35 (87%) | 2.74 (86%) | 3.13 (86%) | 5.07 (83%) | 5.15 (84%) | 5.20 (85%) |
| *Climate model* | *RCP 2.6* | *RCP 4.5* | *RCP 8.5* | *RCP 2.6* | *RCP 4.5* | *RCP 8.5* |

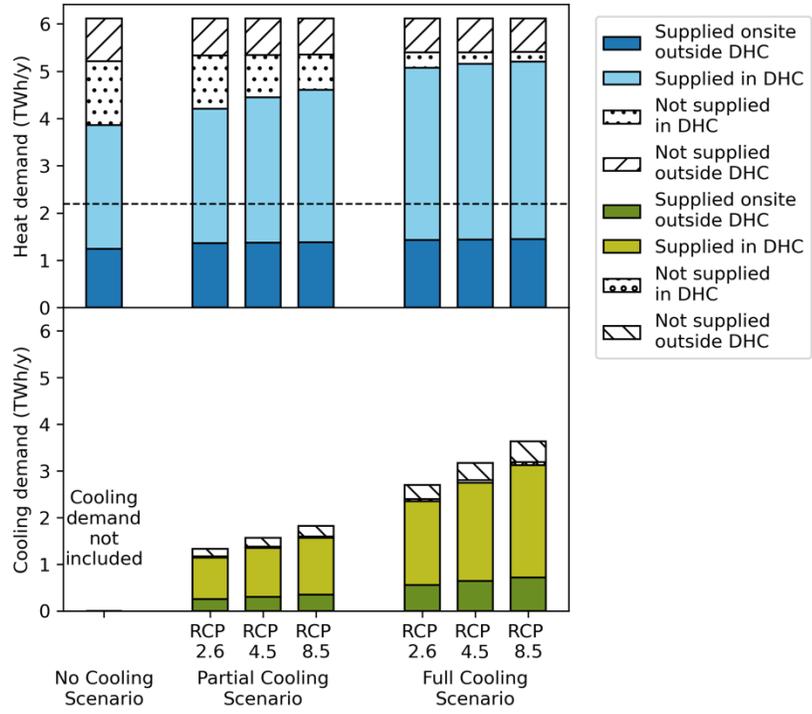

Figure 10. Supply of building heating and cooling demands in scenarios with DHC.

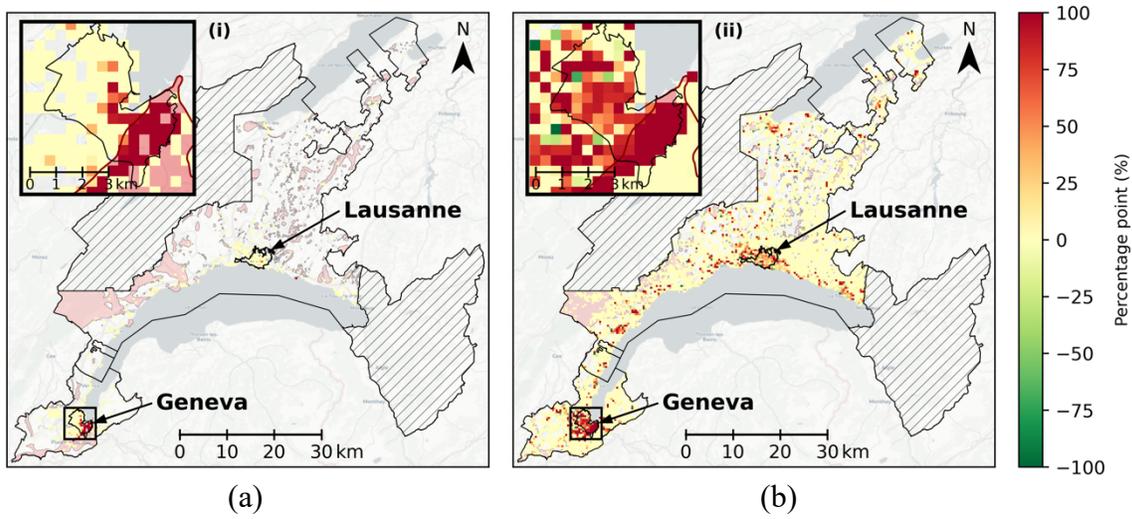

Figure 11. The difference between PC-ND-4.5 and PC-D-4.5 scenarios: a) percentage point difference of cooling demand supplied; b) percentage point difference of heat demand supplied.

# 5 Discussion
## 5.1 Methodological contribution

This paper proposes a novel framework to estimate the technical and useful potential of shallow GSHPs for building heating *and* cooling in individual GSHP systems and in DHC, which scalable to entire regions. For the first time, this framework combines (i) the spatial mapping between heat demands and virtually installed BHEs, (ii) the analytical modelling of the technical heat exchange potential from bi-directional GSHPs, considering seasonal regeneration through the re-injection of space cooling demands, and (iii) the optimal allocation of potential heat sources within DHC. The analytical model, which is built upon previous work on the heat extraction potential from GSHPs [23], to account for (i) technical constraints due to the combined heat injection and extraction, and (ii) different operating strategies of GSHP systems. The advantages of the proposed method are that it (i) accounts for thermal interactions between densely installed GSHPs and the seasonal variation of the energy demand, (ii) proposes a trade-off between operating power and heat exchange potential, and (iii) is scalable to thousands of borehole fields.

We further expanded previous work on the identification of potential DHC areas [52] by introducing a graph-theory based optimization to match building thermal energy demand to technical geothermal potential. This approach permits to quantify the impact of DHC on the useful geothermal potential. The proposed method is transferable to other heat sources for seasonal regeneration, such as solar thermal generation, waste heat or air-source heat pumps [72]. The framework may also be replicated in other regions within or outside of Switzerland. Many of the required regional input data are available at European scale or beyond, or can be mapped from the literature using existing methods. An overview of these datasets and mapping approaches, which assure the replicability beyond the case study, are provided in Appendix B.

## 5.2 Practical implications and application

The results of the case study in the Swiss cantons of Vaud and Geneva imply that seasonal regeneration of GSHP systems may significantly increase the potential heat extraction. We found that the re-injection of space cooling demands into the ground allows for maximum annual heat extraction densities above 300 kWh/m$^2$ at heat injection densities above 330 kWh/m$^2$, especially in centres of large urban areas. The comparison of shallow geothermal potential studies by Bayer et al. [18] indicates that the maximum energy densities correspond to the yields of boreholes with little or no thermal interference [34], [73]. We also found that the maximum technical potential is consistent with a previous Swiss case study for a commercial GSHP installation [46], which suggests a heat extraction density of 440 kWh/m$^2$ at 610kWh/m$^2$ of heat injection. Across the case study region, the heat extraction potential increases with the amount of cooling injection from 4.6 TWh without regeneration up to 6.9 TWh. The spatial resolution of pixels of 400x400m$^2$ highlights regional differences of the shallow geothermal potential between rural, suburban and urban areas.

We further estimate that the conversion rate of injected to additional extracted heat is 90%, showing the high impact of seasonal regeneration on reducing thermal interference. Integrating GSHPs within DHC increases the useful potential to supply building heating demand by 75% without regeneration and by up to 55% with regeneration, allowing to cover up to around 85% of heating and cooling demands through shallow geothermal energy. The fractions are nearly independent of different climate change scenarios for 2050, which implies robustness towards uncertainties in future climate. Low confidence intervals across 2000 Monte Carlo runs also suggest that the spatial distribution of the cooling demand has a low impact on the total potential.

The results may be applied to identify strategic areas for installing GSHPs in individual and district heating systems. This work can further be used as a basis for economic studies to compare the cost-effectiveness of renewable energy sources, therefore helping to design future energy systems. Finally, the findings presented above may provide useful input for policy makers to discuss the regional and national renewable energy strategies.

## 5.3 Limitations and future work

Quantifying the impact of seasonal regeneration of GSHPs is subject to assumptions and limitations related to the data and the modelling approach. The main assumptions and limitations related to data are that (i) all potential GSHP systems and energy demands within each pixel of 400 x 400 m$^2$ are connected. This assumption is necessary to assure the scalability of the approach; (ii) for each potential DHC area, all GSHP systems and buildings are connected to the same DHC.

Any GSHP system adjacent to DHC areas is a potential further heat source; (iii) the mapping of geothermal potential to building energy demands is done by calculating heat balances. DHC network topologies, operating temperatures and thermal losses are not addressed; and (iv) we consider only the space cooling demand in the service sector. Space cooling in the residential sector is not modelled, as the proportion of expected residential space cooling demand for 2050 is small (5%-35%) according to forecasts for neighbouring countries [74].

The key assumptions of the analytical model, which are described in detail in [23], are that (i) the impact of groundwater flow on the technical geothermal potential is neglected, due to a lack of available data. Groundwater flow may impact both the magnitude of the GSHP potential and the direction of thermal interferences, which we assumed to be isotropic in all directions; (ii) the GSHP systems use the full available area and are designed such as to provide the estimated heat extraction rate and annual energy; and (iii) all systems start the heat extraction simultaneously, neglecting any previously existing systems.

Future work is needed to assess the technical barriers for using shallow geothermal energy in future DHC and to account for the impact of groundwater flow on the technical potential. Furthermore, environmental consequences of long-term warming or cooling trends of the ground related to the net injection or extraction of heat may be investigated. To adapt the method to country scale, statistical methods such as Machine Learning may be used. The method proposed here may also be used to model other systems for seasonal regeneration at large scale, such as the re-injection of excess solar thermal generation. This study can be used as the basis for further work to explore the economic and emission reduction potential of shallow geothermal combined with district heating and cooling. Such further work may also address the potential of hybrid GSHPs with other heat sources such as industrial waste heat or solar thermal generation, as well as complementarities with renewable electricity generation from wind turbines or solar photovoltaics.

# 6    Conclusion

This work presents a novel framework to estimate the technical and useful potential of shallow ground-source heat pumps (GSHPs) to supply building heating and cooling demands at regional scale. The framework accounts for the geospatial matching of heat demands and potential GSHP systems, the modelling of technical potential with seasonal regeneration of GSHPs through re-injection of excess heat from space cooling, and for the optimal allocation of heat supply in district heating and cooling (DHC). The useful potential is obtained for the direct heat exchange between buildings and geothermal fields and by considering DHC. A scenario-based approach is used to assess the technical and useful geothermal potential under different climate change scenarios, market penetration levels of cooling systems, and the possible use of DHC.

The case study in western Switzerland suggests that seasonal regeneration may significantly reduce thermal interference between boreholes. This increases the maximum technical geothermal potential density from 15 kWh/m$^2$ without heat injection to above 300 kWh/m$^2$ in pixels with heat injection densities above 330 kWh/m$^2$. These values are consistent with results reported in related scientific literature and case studies of existing installations. Results further suggest that the useful geothermal potential may cover up to 55% of heat demand while covering 57% of service-sector cooling demand in 2050 (assumed at 50% of current heat demand) for individual GSHP systems, which increases to around 85% if DHC is used. The results are robust to uncertainties in future climate. The outcomes of the study may be used to conduct techno-economic analyses of future energy systems with a high share of renewable heat generation and to inform decision-making aimed at achieving Switzerland's renewable energy targets for 2050.

**Appendix A. Heating and cooling degree days**

The heating and cooling degree days (HDD/CDD) are computed from daily mean ambient temperature ($T_{amb}$) [75]. For HDD, we use the formula provided in the Swiss norm SIA 2028 [70]:

$$\text{HDD} = \sum_{d=1}^{d_m} \left(20°C - T_{amb}(d, m)\right) \quad \forall T_{amb}(d, m) \leq 12°C \tag{A.1}$$

where $d, m$ denote the day and month and $d_m$ is the number of days in each month. As no Swiss norm exists for CDD, we obtain the CDD from [75] using a reference temperature of 18°C:

$$CDD = \sum_{d=1}^{d_m} \left(T_{amb}(d, m) - 18°C\right) \quad \forall T_{amb}(d, m) \geq 18°C \tag{A.2}$$

The maximum monthly heating/cooling weights ($w_{hdd/cdd,max}$) are obtained as [23]:

$$w_{hdd/cdd,max} = 1.05 \frac{HDD/CDD_{max}}{\sum HDD/CDD} \tag{A.3}$$

where the $HDD/CDD_{max}$ are the maximum monthly HDD/CDD.

**Appendix B. Data availability beyond the case study area**

To replicate the proposed methods outside of the case study area, the input datasets summarized in Table 1 must be obtained for the area of interest. While many of the required input datasets are available at European scale or beyond, other datasets would have to be approximated. Notably, maps of the thermal ground properties are rarely available. For large-scale studies, these can be mapped from literature data based on geological characteristics, as performed for example in [11], [14], [76], [77]. Such mapping is a crude approximation, and the results must be interpreted in this context. However, related work has shown that the thermal ground properties are not the most important features impacting GSHP performance [78], so this approach may be acceptable for large-scale studies.

Furthermore, the quality and resolution of the input data have to be suitable for the target application. The proposed analytical model is applied at individual parcel scale, which can be derived for example from OpenStreetMap data. For analyses at country scale or at pan-European scale, for example, even low-resolution inputs, for example using heat demand pixels at km-scale, may be acceptable for the replication of the proposed framework. At city scale, data is often available at higher quality and higher spatial resolution, allowing to obtain more accurate results. An overview of the required input datasets, there potential availability in the European context and suggestions of references are provided in Table 8.

Table 8. Suggestions of methods and datasets to obtain the required input data for the replicability of the proposed framework beyond the case study region. References are provided for the European scale where possible, but they are not exhaustive.

|  | Input dataset | Potential availability | References (Europe) |
|---|---|---|---|
| Spatial constraints | Available area for BHEs | Can be obtained from national topographic data or OpenStreetMap (OSM) following the method in [23] | OSM [79] |
|  | DHC zones | Can be obtained from the Heat Roadmap Europe [1] or derived using the method proposed in [67] | Heat Roadmap Europe [80] |
|  | GSHP restrictions | Can be neglected for a rough potential estimate, or estimated from national hydrogeological data | SwissTopo [81] (Switzerland) |
| Geothermal input data | Thermal ground properties | If no 3D underground models or measurements are available, mapping of literature values to geological data may be performed in analogy to related studies [11], [14], [76], [77] | SIA [45], VDI [82] |
|  | Surface temperature | Can be derived from air temperature (see below) following [45], [83] | COSMO [2] |
|  | Air temperature | Reanalysis (e.g. COSMO REA) or other gridded data can be used |  |
|  | Full-load heating hours | Can be obtained from literature/norms | SIA [45], Kavanaugh [48] |
| Building energy demands | Building heating demand | Can be obtained from the Heat Roadmap Europe[1] for the reference year 2015. Future penetration of cooling equipment must be assessed in further work. | Heat Roadmap Europe [84], Hotmaps [85] |
|  | Building cooling demand |  | Heat Roadmap Europe [86] |

---

[1] https://heatroadmap.eu/peta4/

[2] https://reanalysis.meteo.uni-bonn.de/

**Research data for this article**

The research data related to this article will shortly be made available at:
https://doi.org/10.5281/zenodo.5575318

**Acknowledgements**

We thank the reviewers and the editor for thorough reviews and helpful comments on an earlier version of this article. This research has been financed partly by the Swiss Innovation Agency Innosuisse under the Swiss Competence Center for Energy Research SCCER FEEB&D and partly by the Swiss National Science Foundation (SNSF) under the National Research Program 75 (Big Data).

**Author contribution statement**

**Alina Walch & Xiang Li:** Conceptualization, Methodology, Software, Visualization, Writing - Original draft. **Jonathan Chambers** Conceptualization, Methodology, Writing - Review & Editing, Supervision. **Selin Yilmaz:** Writing - Review & Editing, Supervision. **Nahid Mohajeri:** Conceptualization, Writing - Review & Editing, Supervision, Funding acquisition. **Jean-Louis Scartezzini & Martin Patel:** Supervision, Project administration, Funding acquisition.
34